\documentclass[aps,prl,reprint,letterpaper,amsmath,amssymb, superscriptaddress,longbibliography]{revtex4-1}
\usepackage{amsmath}
\usepackage{color}
\usepackage{graphicx}

\begin{document}

\title{Optical detection of electron spin dynamics driven by fast variations of a magnetic field: a simple method to measure $T_1$, $T_2$, and $T_2^*$ in semiconductors}

\author{V.~V.~Belykh}
\email[]{belykh@lebedev.ru}
\affiliation{P.N. Lebedev Physical Institute of the Russian Academy of Sciences, 119991 Moscow, Russia}
\author{D.~R.~Yakovlev}
\affiliation{Experimentelle Physik 2, Technische Universit\"{a}t Dortmund, D-44221 Dortmund, Germany}
\affiliation{Ioffe Institute, Russian Academy of Sciences, 194021 St. Petersburg, Russia}
\author{M.~Bayer}
\affiliation{Experimentelle Physik 2, Technische Universit\"{a}t Dortmund, D-44221 Dortmund, Germany}
\affiliation{Ioffe Institute, Russian Academy of Sciences, 194021 St. Petersburg, Russia}


\begin{abstract}
We develop a simple method for measuring the electron spin relaxation times $T_1$, $T_2$ and $T_2^*$ in semiconductors and demonstrate its exemplary application to $n$-type GaAs. Using an abrupt variation of the magnetic field acting on electron spins, we detect the spin evolution by measuring the Faraday rotation of a short laser pulse. Depending on the magnetic field orientation, this allows us to measure either the longitudinal spin relaxation time $T_1$ or the inhomogeneous transverse spin dephasing time $T_2^*$. In order to determine the homogeneous spin coherence time $T_2$, we apply a pulse of an oscillating radiofrequency (rf) field resonant with the Larmor frequency and detect the subsequent decay of the spin precession. The amplitude of the rf-driven spin precession is significantly enhanced upon additional optical pumping along the magnetic field.
\end{abstract}

\maketitle
%
%
\thispagestyle{empty}

\section*{Introduction}

The electron spin dynamics in semiconductors can be addressed by a number of versatile methods involving electromagnetic radiation either in the optical range, resonant with intraband transitions, or in the microwave as well as radiofrequency (rf) range, resonant with the Zeeman splitting \cite{Dyakonov2017,Goldfarb2018}. For a long time, the optical methods were mostly represented by Hanle effect measurements giving access to the spin relaxation time at zero magnetic field \cite{Meier1984}. New techniques giving access both to the spin $g$ factor and relaxation times at arbitrary magnetic fields have been very actively developed. These are resonant spin amplification \cite{Kikkawa1998}, with its single-beam modification \cite{Saeed2018}, and spin noise spectroscopy \cite{Aleksandrov1981,Crooker2004,Oestreich2005}, which was recently improved to increase its sensitivity,  extend the accessible range of magnetic fields \cite{Cronenberger2016,Petrov2018,Muller2010}, enable determination of the homogeneous spin relaxation time $T_2$ \cite{Yang2014,Braun2007,Poshakinskiy2019}, and even make it sensitive to spin diffusion \cite{Cronenberger2019}. Nevertheless, the most powerful optical technique is pump-probe Faraday/Kerr rotation \cite{Baumberg1994,Zheludev1994,Awschalom2002,Yakovlev2017}, which through the interband electron transitions allows one to directly address the dynamics of a spin ensemble with (sub)picosecond resolution in a broad temporal range \cite{Belykh2016,Belykh2017}. The non-optical methods are mostly represented by electron paramagnetic resonance (EPR) spectroscopy where a microwave (or rf) field resonant with the Larmor frequency of the electron spin precession is absorbed by the sample. This method can be used also in the time-resolved mode (pulsed EPR) \cite{Blume1958,Gordon1958,Schweiger2001}. Recently, a combination of the pump-probe and EPR techniques was proposed, where the electron spin ensemble is driven by the rf pump pulse and the spin precession is probed optically by a laser pulse, for which the Faraday rotation is measured \cite{Belykh2019}. This radio-optical pump-probe technique is easy in implementation and resonantly addresses spin states allowing to measure the homogeneous spin relaxation time $T_2$.

In this paper, we make the next step forward in the development of the radio-optical pump-probe technique: (i) We show that the amplitude and direction of the spin precession that is driven by the rf field can be controlled by additional circularly polarized optical pumping along the applied constant magnetic field. Thereby, the signal level can be significantly increased to make the technique applicable for $T_2$ determination of numerous material systems. (ii) We introduce a scheme for measuring also the longitudinal spin relaxation time $T_1$ and inhomogeneous transverse spin dephasing time $T_2^*$. We exploit the fact that the equilibrium spin polarization is controlled by the magnetic field, and create an abrupt, small change in the magnetic field magnitude or direction with an rf coil, in order to observe the spin evolution towards a new equilibrium state.

\section*{Details of the technique}
\begin{figure*}
\includegraphics[width=1.6\columnwidth]{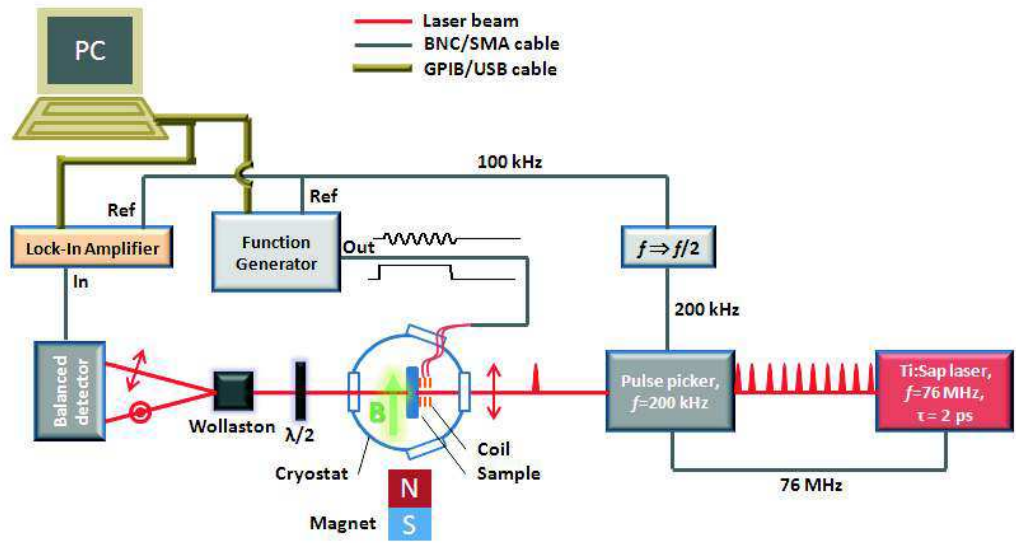}
\caption{Scheme of the experiment. A rapid change of the magnetic field produced by the coil connected to the function generator induces the electron spin polarization dynamics in the sample. The spin dynamics is probed by the Faraday rotation of the linear polarization of the optical pulses picked from a train of laser pulses. The dynamics of spin precession is scanned by changing the delay between the synchronized RF and optical pulses. Note that the constant magnetic field may be applied either in the Voigt (shown) or the Faraday geometry.}
\label{fig:Setup}
\end{figure*}

The detailed experimental scheme is shown in Fig.~\ref{fig:Setup}. The experiment is performed on a Si-doped GaAs sample with the electron concentration of $1.4 \times 10^{16}$~cm$^{-3}$ (350-$\mu$m-thick bulk wafer) near the metal-to-insulator transition \cite{Belykh2016,Belykh2019}. The sample is placed in a variable temperature He-bath cryostat ($T =1.6-300$~K) with a permanent magnet placed outside on a controllable distance. A constant magnetic field $\mathbf{B}$ up to 15~mT is applied either perpendicular or parallel to the light propagation direction and to the sample normal (Voigt and Faraday geometry, respectively). In the high-frequency experiments the sample is placed in a cryostat with a split-coil superconducting magnet for generating magnetic fields up to 6~T. The rf magnetic field $\mathbf{b}(t)$ with an amplitude up to 2~mT is applied along the sample normal using a small coil ($\lesssim 1$~mm-inner and $\sim 1.5$~mm-outer diameter), placed near the sample surface. The current through the coil is driven by a function generator (except for the high-frequency experiments described in the last section), which creates voltage pulses of either sinusoidal [up to 75~MHz, Fig.~\ref{fig:T1T2}(e)] or rectangular [Fig.~\ref{fig:T1T2}(a) and \ref{fig:T1T2}(c)] form. The rf field pulse created by the coil acts as pump pulse in the spin dynamics experiment. For probing, we use 2-ps-long optical pulses generated by a Ti:Sapphire laser. The laser emits a train of pulses with a repetition rate of 76~MHz, which is reduced to 200~kHz (exemplary value used in our experiment) by selecting single pulses with an acousto-optical pulse picker that is synchronized with the laser. The arrival of the rf pulses is synchronized with the laser pulses using a signal from the pulse picker that triggers the function generator. The repetition frequency of the rf pulses is reduced by a factor of 2 with respect to that of the laser pulses. The latter is required for synchronous detection at the repetition frequency of the rf pulses of 100~kHz. Thus, the difference of the optical signal, with and without rf pulse is detected. The delay between the pump rf pulses and the probe laser pulses is varied electronically using a function generator, see also Ref.\cite{Belykh2019}. The probe laser pulses are linearly polarized and their polarization after transmission through the sample is rotated by 45$^\circ$ with respect to the horizontal plane using a $\lambda/2$ wave plate. In what follows, using a Wollaston prism, the  probe beam is split into two orthogonally polarized beams of approximately equal intensities which are registered by a Si-based balanced photodetector. The Faraday rotation of the probe beam polarization induced in the sample is detected as an imbalance in the intensities of the above-mentioned two beams. The laser wavelength is set to 827~nm. 

In the experiments with additional optical pumping [Fig.~\ref{fig:OP}~(b)], the center of the rf coil and probe spot are close to the sample edge. This edge is illuminated from the side, almost perpendicularly to the sample normal, by the laser beam focused onto an 0.5~mm spot to overlap with the probe beam inside the sample. The additional illumination is split off from the emission of the same pulsed Ti:Sapphire laser before the pulse picker and modulated with an electro-optical modulator (EOM) to produce 1-$\mu$s-long pulses synchronized with the rf pulses. The characteristic rise and decay times of the pulses behind the EOM are about 200~ns. The helicity of the optical pumping is set with the linear polarizer and $\lambda/4$ wave plate.

\section*{Measurement of $T_1$, $T_2^*$ and $T_2$}
\begin{figure*}
\includegraphics[width=1.9\columnwidth]{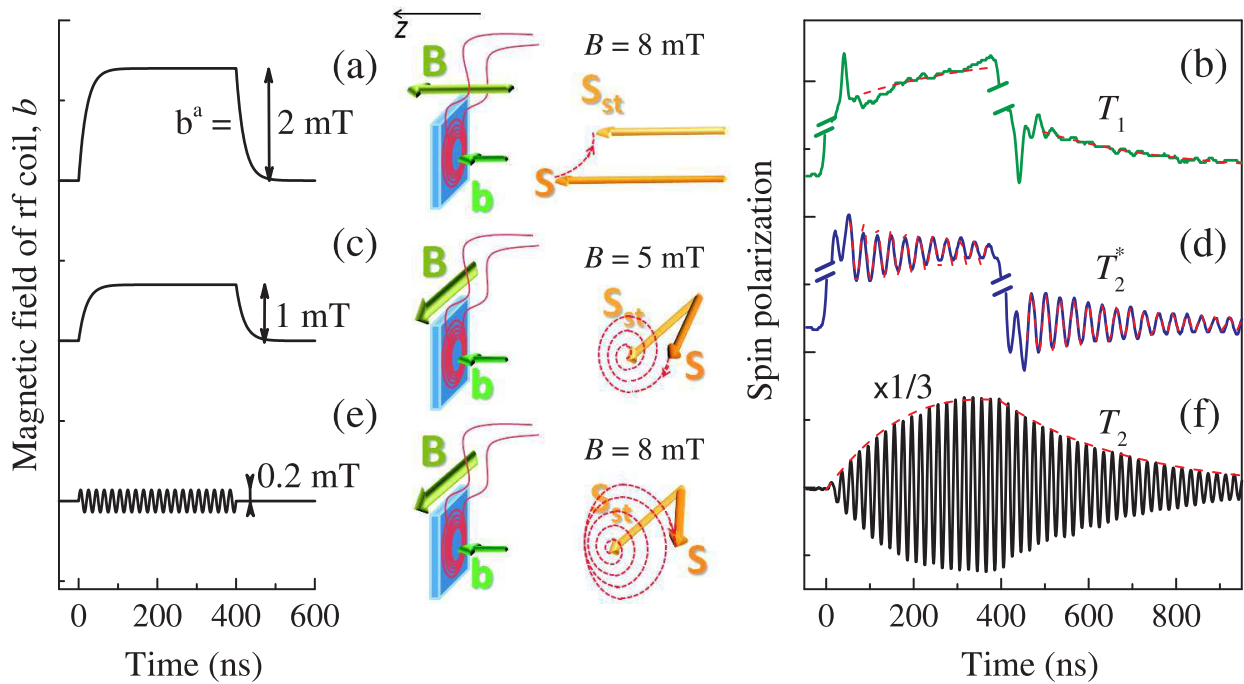}
\caption{Measurement of $T_1$, $T_2^*$ and $T_2$. From left to right: temporal profiles of the rf coil field $b(t)$, schematic presentations of the experimental geometries, evolutions of the spin polarization, and corresponding measured dynamics of the electron spin polarization along the $z$-axis. (a),(b) The constant field $\mathbf{B}$ in the Faraday geometry and step-like profile of $b(t)$ allow one to measure $T_1$. (c),(d) The $\mathbf{B}$ in the Voigt geometry and step-like profile of $b(t)$ allow one to measure $T_2^*$. (e),(f) The $\mathbf{B}$ in the Voigt geometry and sinusoidal profile of $b(t)$ resonant with the Larmor frequency [$\omega=\omega_\text{L}=2\pi \times$(50~MHz)] allow one to measure $T_2$. The red dashed lines in Figs.~\ref{fig:T1T2}(b),(d),(f) show model fits to the experimental data. $T = 2$~K.}
\label{fig:T1T2}
\end{figure*}

\begin{figure}
\includegraphics[width=\columnwidth]{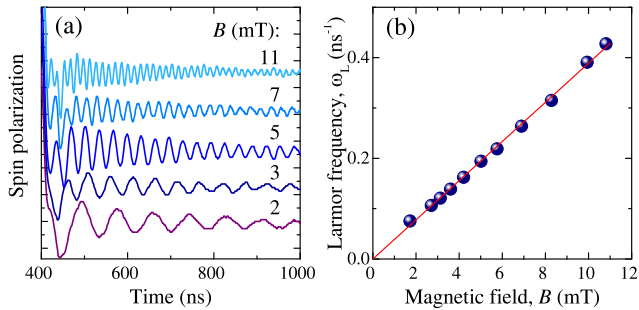}
\caption{Magnetic field dependence of spin precession. (a) Dynamics of the electron spin polarization after the step-like change of $b(t)$ [configuration of Fig.~\ref{fig:T1T2}(c)] for different values of the constant magnetic field $\mathbf{B}$ applied in the Voigt geometry. (b) Larmor precession frequency $\omega_\text{L}$ as a function of the magnetic field $B$. $T = 2$~K.}
\label{fig:T2Star}
\end{figure}

\begin{table*}
\begin{center}
\begin{tabular}{| c | c | c | c |}
    \hline
    \textbf{Experimental configuration} & \textbf{Measured spin} & \textbf{Spin polarization} & \textbf{Limitations}\\
     & \textbf{relaxation time} & \textbf{amplitude} & \\
    \hline
    Faraday $\mathbf{B} \parallel \mathbf{b}$, step-like $b(t)$ & $T_1$ & $\chi b$ & Level of the signal\\
    \hline
    Voigt $\mathbf{B} \perp \mathbf{b}$, step-like $b(t)$ & $T_2^*$ & $\chi b$ & Level of the signal, magnetic field\\
    \hline
    Voigt $\mathbf{B} \perp \mathbf{b}$, sinusoidal $b(t)$ & $T_2$ & $\displaystyle \frac{1}{2}\frac{T_2^*}{\sqrt{T_1 T_2}}\chi B$ & None\\
    \hline
    \end{tabular}

    \caption{Summary of the experimental configurations used to measure the different spin relaxation times.}
\label{Tab}
\end{center}
\end{table*}

We study the spin dynamics of resident electrons in the conduction band of a bulk GaAs sample which were introduced by doping. In the external magnetic field $\mathbf{B}$, the electron ensemble has an equilibrium spin polarization $\mathbf{S}$ which is determined by thermal distribution between Zeeman-split spin levels. Spin polarization $\mathbf{S}$ is proportional to and directed along $\mathbf{B}$. Using this fact, we rapidly change $\mathbf{B}$ and monitor the evolution of the spin polarization towards a new equilibrium state.

In the first experiment [Fig.~\ref{fig:T1T2}(a)], the magnetic field $\mathbf{B}$ is applied in the Faraday geometry along the sample normal ($z$-axis). Using the rf coil, we induce a small ($\lesssim 2$~mT) and rapid ($\sim 20$~ns) change in the magnetic field magnitude by applying the additional magnetic field $\mathbf{b}(t) \parallel \mathbf{B}$ with a step-like temporal profile. Just after the beginning of the $b(t)$ pulse and after its end we observe the longitudinal spin relaxation towards a new equilibrium state occurring on the time scale of $T_1\approx 300$~ns [Fig.~\ref{fig:T1T2}(b)]. The signal is also contributed by the Faraday rotation that is inevitably present in bulk GaAs and not related to the spin polarization of resident carriers. This contribution instantaneously follows the $b(t)$ profile and does not contribute to the decaying part of the dynamics. The short spikes at the beginning and at the end of the $b(t)$ pulse are caused by transient processes in the rf contour.

In the second experiment, which is also based on the repolarization of the electron spin system, $\mathbf{B}$ is applied in the Voigt geometry perpendicular to the sample normal [Fig.~\ref{fig:T1T2}(c)]. The rapid application of the field $\mathbf{b} \perp \mathbf{B}$ using the rf coil induces a rapid tilt of the total magnetic field $\mathbf{B}_\text{tot} = \mathbf{B} + \mathbf{b}(t)$ by an angle $\approx b/B$. As a consequence, we observe spin precession about the new direction of the magnetic field with the Larmor frequency $\omega_\text{L} = |g|\mu_\text{B}B_\text{tot}/\hslash$ and relaxation with the inhomogeneous spin dephasing time $T_2^*\approx 300$~ns [Fig.~\ref{fig:T1T2}(d)]. The dynamics of the spin polarization for different values of $B$ in this experiment are shown in Fig.~\ref{fig:T2Star}(a). The dependence of the spin precession frequency $\omega_\text{L}$ on $B$ is shown in Fig.~\ref{fig:T2Star}(b). As expected, $\omega_\text{L}$ increases linearly with $B$ allowing to determine $|g| = 0.44$. It is known that in GaAs $g < 0$, so $g = -0.44$. 

In the third experiment [Fig.~\ref{fig:T1T2}(e)], $\mathbf{B}$ is still oriented in the Voigt geometry and we apply the oscillating rf field $\mathbf{b}(t) = \mathbf{b}^a\sin(\omega t)$ with $\omega = \omega_\text{L}$, which causes a gradual declination of the spin polarization $\mathbf{S}$ from the direction of $\mathbf{B}$ and its precession around $\mathbf{B}$ [Fig.~\ref{fig:T1T2}(f)], i.e. the onset of a transverse oscillating component of $\mathbf{S}$. Note that in an electron spin ensemble with inhomogeneous $g$-factor distribution, the rf field affects only the spins with $\omega_\text{L} - 1/T_2 \lesssim \omega \lesssim \omega_\text{L} + 1/T_2$ \cite{Belykh2019} unlike in the previous experiment, where all spins become involved. Thus, after the rf field is switched off, only these spins contribute to the spin polarization decay with a decay time close to $T_2$ rather than $T_2^*$. From the experiment we obtain the spin precession decay time $T_2 \approx 300$~ns. In the studied $n$-GaAs sample the electron density exceeds the metal-insulator transition and the majority of the electrons are free. Therefore, the individual spin properties are shared in the spin ensemble due to spin exchange averaging \cite{Pines1955,Paget1981}, leading to $T_2 \approx T_2^*$. For the samples with strongly localized electrons one would expect $T_2^* \ll T_2$ especially at high $B$.

Thus, these three experimental configurations allow us to measure all spin relaxation times $T_1$, $T_2$, and $T_2^*$. Nevertheless, the applicability of the radio-optical pump-probe technique to a wide range of systems depends on the signal level as we discuss below. This technique is based on inducing a controllable deviation of the spin polarization from its equilibrium (stationary) value $\mathbf{S}_\text{st}$ which is determined by the thermal population of Zeeman levels. More specifically \cite{Belykh2019}:
\begin{equation}
\mathbf{S}_\text{st} = \chi \mathbf{B}_\text{tot},
\end{equation}
\begin{multline}
\chi \approx -\frac{g\mu_\text{B}}{12 n_0}\frac{\partial n(\mu, T)}{\partial \mu}|_{\mu = \mu_0} \\
\approx
-g\mu_\text{B} \times
\begin{cases}
   \frac{1}{12 k_\text{B} T}, & k_\text{B} T \gg \epsilon_\text{F}\\
   \frac{1}{8 \epsilon_\text{F}}, & k_\text{B} T \ll \epsilon_\text{F}.
 \end{cases}
\end{multline}
Here we assume that the maximal spin polarization is 1/2, $\mu_\text{B}$ is the Bohr magneton, $k_\text{B}$ is the Boltzmann constant, $\mu$ is the chemical potential (in the limit of $T=0$ it is identical to the Fermi energy $\epsilon_\text{F}$), $n_0 = n(\mu_0, T)$ is the total electron concentration, and we take into account that $|g|\mu_\text{B}B \ll k_\text{B} T, \epsilon_\text{F}$. Note, $\chi > 0$ for $g < 0$.

In the first two experimental configurations, after the rapid change of the magnetic field, the spin polarization deviates from its new equilibrium value by $\chi b$ which determines the amplitude of the signal. In the third (EPR-based) experimental configuration a significant fraction of the initial equilibrium spin polarization $\chi B$ is involved in the spin precession and its amplitude reaches the maximal value $S_z^\text{max} = \chi B T_2^* / 2 \sqrt{T_1 T_2}$ for the rf field amplitude $b^a = 2\hslash/|g|\mu_\text{B} \sqrt{T_1 T_2}$ (see the Model section). These results are summarized in Table~\ref{Tab}. The dynamics of the spin polarization in all three experimental configurations are calculated in the Model section below and reproduce well the experimental observations [model fits to the experimental data are shown by the red dashed lines in Figs.~\ref{fig:T1T2}(b),(d),(f)].

\section*{Signal enhancement with additional optical pumping}
From the above discussion and Table~\ref{Tab} one can see that in the repolarization (the first two) experiments, the level of the signal is almost independent on $B$ and proportional to the rf coil field $b$ which is difficult to make larger than a few milliTesla. On the other hand, in the EPR (third) experiment, the signal quickly saturates with $b$ (to be precise it shows Rabi oscillations as a function of $b$) and is proportional to the magnetic field $B$, which can be as large as several Tesla. Correspondingly, the rf field frequency should be tuned towards the GHz range to be resonant with the electron Zeeman splitting.

\begin{figure*}
\includegraphics[width=1.8\columnwidth]{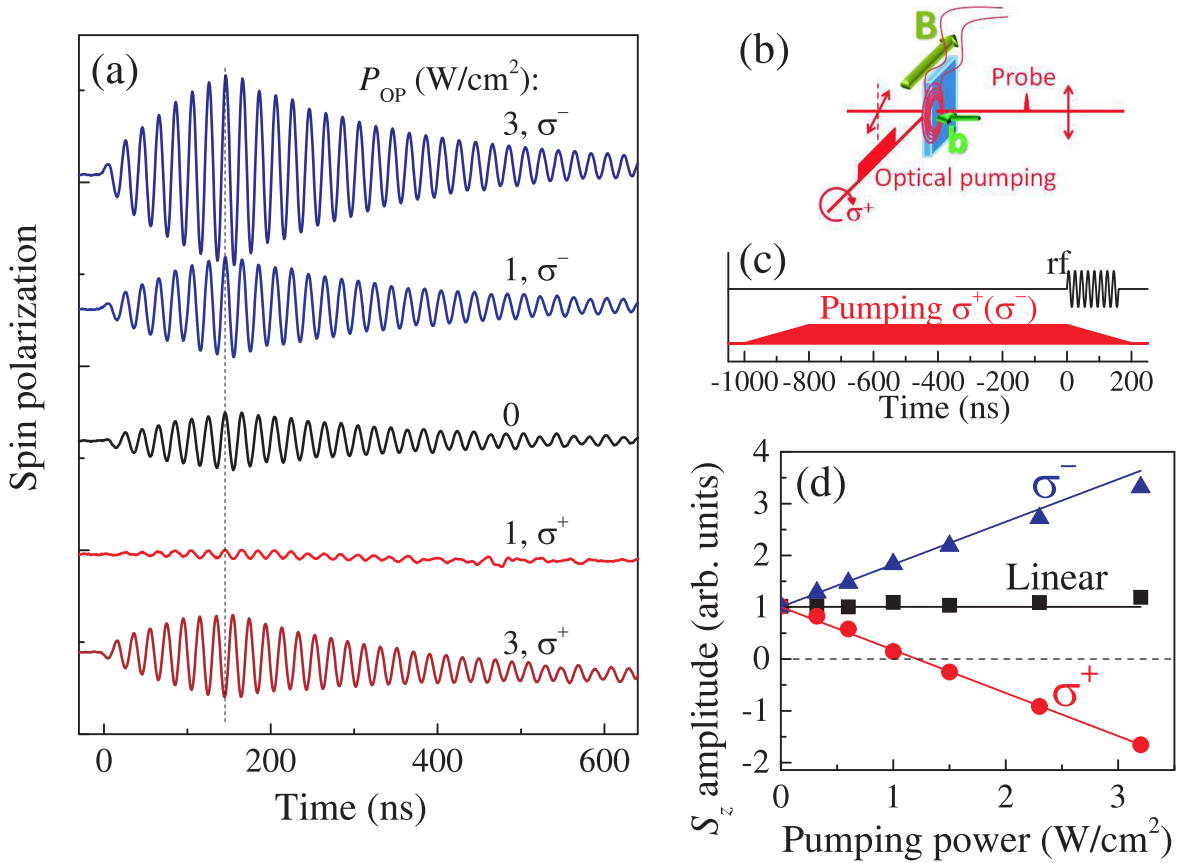}
\caption{Effect of optical pumping. (a) Dynamics of the electron spin polarization created by the rf field resonant with the Larmor frequency for different powers and helicities of the optical pumping incident on the sample edge. $B = 8.3$~mT, $\omega=\omega_\text{L}=2\pi \times$(50~MHz), $b^\text{a} = 0.2$~mT, rf pulse duration is 160~ns. (b) Scheme of the experiment with optical pumping. (c) Temporal profiles of the rf field $b$ and optical pump. (d) Spin precession amplitude just at the end of the rf pulse as a function of the optical pump power density for different helicities. Lines show linear fits. $T = 6$~K.}
\label{fig:OP}
\end{figure*}
There is another way to increase the signal in the EPR-based experiment [Fig.~\ref{fig:T1T2}(e),(f)]. In general, the level of the signal in EPR is determined by the spin polarization along $\mathbf{B}$ which is usually in equilibrium, $S_\text{st}=\chi B$. However, we can create nonequilibrium spin polarization by illuminating the sample edge along $\mathbf{B}$ (perpendicular to the sample normal) with circularly polarized laser light \cite{Meier1984}. The advantage of this optical pumping has been already realized in the continuous wave EPR experiments \cite{Hermann1971,Cavenett1981}. Additional pumping has a dual effect on the spin precession induced by the rf field. First, it depolarizes the precessing spin component and reduces its amplitude and dephasing time \cite{Heisterkamp2015}. The second effect of the optical pumping is the desired one: It generates a nonequilibrium spin polarization $S_\text{OP} \propto P_\text{OP}$ ($P_\text{OP}$ is the pump power density) which contributes to the equilibrium polarization $S_\text{st}=\chi B$. The scheme of this experiment is given in Fig.~\ref{fig:OP}(b), and the temporal profiles of the rf coil field and optical pumping are shown in Fig.~\ref{fig:OP}(c). We switch the optical pump off just at the onset of the rf pulse, so it almost does not perturb the spin precession, while the nonequilibrium spin polarization, that it has generated, $S_\text{OP}$, is still present (it decays with time $T_1$). The other advantage of the limited duration of the additional illumination is reduction of nuclear spin polarization. Electron spins, oriented by the additional illumination, polarize the nuclear spins by acting on them by the effective Knight magnetic field \cite{Knight1949}. A nuclear spin polarization, in turn, creates an effective Overhauser magnetic field acting back on the electron spins \cite{Overhauser1953}. This leads to a shift of the electron spin Larmor frequency with sign and magnitude dependent on the helicity and intensity of the pumping, respectively, which is unwanted in our experiments. 

Figure~\ref{fig:OP}(a) shows the electron spin polarization dynamics during the rf pulse having the frequency resonant with $\omega_\text{L}$ for different power densities and helicities of the optical pump. Interestingly, additional pumping with $\sigma^-$ helicity enhances the spin polarization precession amplitude, while increasing power of $\sigma^+$ pumping first completely suppresses the spin precession signal before enhancing it. When crossing the zero with increasing $\sigma^+$ pumping density, the spin polarization oscillations change phase by $\pi$ [see the dashed line in Fig.~\ref{fig:OP}(a)], i.e. the spin precession reverses its direction. The dependence of the spin precession amplitude just after the rf pulse on the optical pump power density is summarized in Fig.~\ref{fig:OP}(d). For $\sigma^-$ pumping (triangles) it increases linearly by more than a factor of three. For $\sigma^+$ pumping (circles) the amplitude decreases linearly with the opposite slope with respect to the $\sigma^-$ pumping case. Here we assume that the amplitude becomes negative when the precession flips the direction. And finally, we have checked that for linearly polarized pumping (squares), which produces no spin polarization, the spin precession amplitude is unchanged.

This behavior can be easily explained. The total spin polarization along $\mathbf{B}$ for $\sigma\pm$ pumping is $S_\text{st} + S_\text{OP}^{\sigma\pm}$, where $S_\text{OP}^{\sigma\pm}\propto P_\text{OP}$. Correspondingly the spin precession amplitude after the rf pulse (see the Model section) is given by:
\begin{equation}
S_{\sigma\pm}^a = S^\text{a} \cdot (1 \mp \kappa P_\text{OP}),
\label{eq:SP}
\end{equation}
where $S^\text{a}$ is the spin precession amplitude without optical pumping and $\kappa$ is a coefficient dependent on the spin orientation efficiency with optical pumping. Interestingly, for $n$-GaAs in the metallic phase, like the sample under study, the optical pumping efficiency is extremely low especially for photon energies below the band gap \cite{Dzhioev2002}. Nevertheless, we can significantly enhance the spin precession amplitude by applying additional circularly polarized pumping. For systems with localized electrons, where optical pumping is efficient, we expect a strong enhancement of the spin signal. Note, this experiment provides access to the homogeneous spin relaxation time $T_2$, unlike all-optical pump-probe techniques.

\section*{Towards high-frequency experiments}
\begin{figure*}
\includegraphics[width=2\columnwidth]{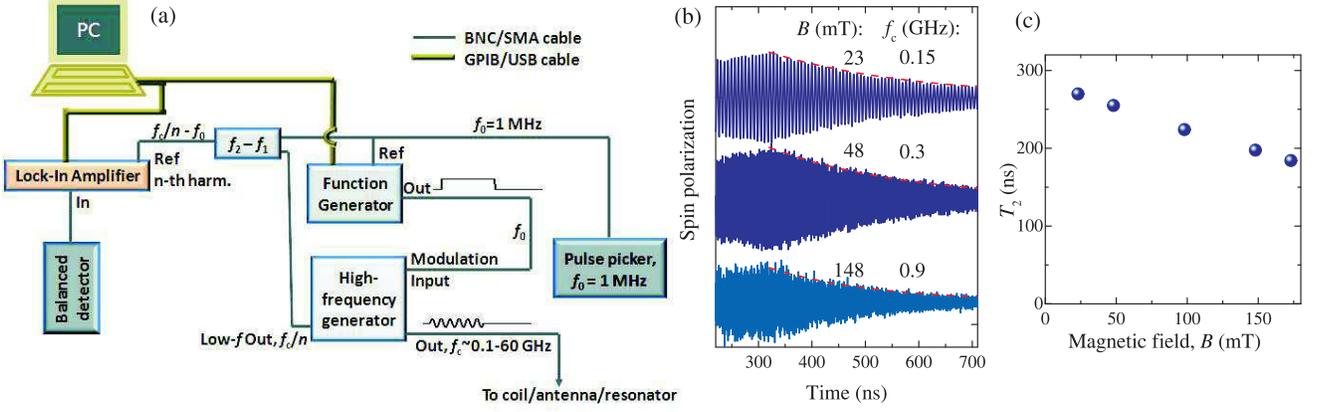}
\caption{High-frequency experiments. (a) Modified electronic part of the experimental scheme, allowing for high-frequency experiments. The rf bursts are produced by the high-frequency generator, while their delay and duration are controlled by the low-frequency function generator synchronized with the laser pulses. (b) Dynamics of the electron spin polarization in the extended range of constant magnetic fields $B$ applied in the Voigt geometry, corresponding to the rf frequencies up to 1~GHz. The red dashed lines show exponential decays. (c) Spin coherence time $T_2$ as a function of magnetic field $B$. The duration of the rf bursts is 320~ns, $T = 2$~K.}
\label{fig:HighF}
\end{figure*}
The key point in the realization of the EPR-like pump-probe experiments [Fig.~\ref{fig:T1T2}(e),(f)] is the phase of the sinusoidal voltage profile which should be synchronized with the laser pulses and with the start of the burst. Thus, in each burst, the rf oscillations should start from the same phase. This goal is easily achieved with an arbitrary function generator producing the required voltage profiles synchronous with the laser pulses. However, the frequency range of such voltage generators is limited, which limits the magnetic field at which spin dynamics can be measured. High-frequency generators, on the other hand, can produce sinusoidal rf or microwave output which can be modulated either internally or externally to produce bursts. In the case of internal modulation, the rf phase may be fixed within each burst, but it is hardly possible to synchronize the bursts with the laser pulses. When the bursts are triggered externally by the signal from the laser (pulse picker) and, thus, synchronized with the laser pulses, the rf phase varies from burst to burst, smashing the signal in the experiment. We show that in the latter case the spin dynamics nevertheless can be measured with a modified electronic part of the setup [Fig.~\ref{fig:HighF}(a)].

The trigger from the laser pulses selected by the pulse picker at frequency $f_0 \approx 1$~MHz (we give here as an example the frequency values used in our experiments) is transformed by the arbitrary function generator to a rectangular pulse of the desired length. The pulse is sent to the high-frequency generator and modulated by the carrier frequency $f_\text{c} = 0.1-67$~GHz, forming the burst that is applied to the coil/antenna/resonator near the sample. Note, $2\pi f_\text{c}$ should be close to the Larmor frequency $\omega_\text{L}$. If we consider the $k$-th laser pulse delayed from the rf burst by $\Delta t$ (the delay is controlled by the function generator), the spin polarization at the moment of the pulse is $S = A\exp(-\Delta t/T_2)\cos(\omega_\text{L}\Delta t + 2 \pi k f_\text{c} / f_0)$. The phase term reflects the phase variation from pulse to pulse. We can select the carrier frequency $f_\text{c} = n f_0 + \Delta f$, where $n$ is an integer and $\Delta f$ is a small enough frequency so that the frequency $n \Delta f$ is in the operation range of the lock-in amplifier. Thus, the balanced detector registers a signal proportional to $S(t) = A\exp(-\Delta t/T_2)\cos(\omega_\text{L}\Delta t + 2 \pi \Delta f t)$, varying as a function of the time $t = k/f_0$. This signal is demodulated by the lock-in amplifier giving the signals $A \exp(-\Delta t/T_2)\cos(\omega_\text{L}\Delta t)$ and $-A\exp(-\Delta t/T_2)\sin(\omega_\text{L}\Delta t)$ in the $X$ and $Y$ channels, respectively. To provide the reference for the lock-in amplifier at frequency $\Delta f$, we take the signal at frequency $f_\text{c} / n$ (which is synchronous with the signal on $f_\text{c}$) from the low-frequency output of the generator, mix it with the signal at $f_0$ and using the low-frequency filter select the component with the difference frequency $f_\text{c} / n - f_0$. This component is received as reference by the lock-in amplifier which uses its $n$-th harmonic for synchronization. Thus, this scheme enables spin dynamics excitation by the rf/microwave radiation in the GHz frequency range which is then measured through optical detection. Here the jitter between the rf bursts and laser pulses impose the main limitation on the maximal Larmor frequency (and magnetic field). This jitter can be as small as few tens of ps corresponding to a magnetic field of several Tesla.

The dynamics of the spin precession measured by this technique up to $f_\text{c} \sim 1$~GHz is shown in Fig.~\ref{fig:HighF}(b). At higher frequencies the rf coil used to deliver the oscillating magnetic field to the sample becomes extremely inefficient and should be replaced by a microwave antenna or resonator for specific frequencies. The dependence of the spin coherence time $T_2$ on the magnetic field is shown in Fig.~\ref{fig:HighF}(c). $T_2$ decreases with $B$ similar to the inhomogeneous spin dephasing time $T_2^*$ \cite{Belykh2016}. The decrease of $T_2^*$ with $B$ is usually attributed to the spread of electron $g$ factors $\delta g$ leading to a corresponding spread of Larmor frequencies, so that, $1/T_2^*(B) = 1/T_2^*(0)+\delta g \mu_\text{B} B/\hslash$. In the experiment in Fig.~\ref{fig:HighF} the rf field affects an electron subensemble with a more narrow distribution of Larmor frequencies than one may expect from $\delta g$. In the case of localized noninteracting electrons we would expect a suppressed decrease of $T_2$ with $B$. However, in our case the electron density is around the metal-insulator transition, so that the electrons are delocalized and efficient spin exchange averaging is present \cite{Pines1955,Paget1981}. As a result, the subensemble affected by the rf share its spin polarization with the entire electron ensemble on a picosecond time scale, leading to the recovery of the Larmor frequency distribution. Thus, for a system with delocalized electrons $T_2$ and $T_2^*$ are hardly distinguishable.

\section*{Model}
\label{App}
Here we show that depending on the orientation of the constant magnetic field and on the temporal profile of the rf field it is possible to measure longitudinal and transverse inhomogeneous as well as homogeneous spin relaxation.
The evolution of the electron spin polarization $\mathbf{S}$ in a magnetic field is described by the Bloch equation \cite{Bloch1946,Abragam1961}:
\begin{equation}
\frac{d\mathbf{S}}{dt} = -\frac{g\mu_\text{B}}{\hslash} \mathbf{S} \times \mathbf{B}_\text{tot} - \hat{\gamma}(\mathbf{S} - \mathbf{S}_\text{st}),
\label{eq:Bloch}
\end{equation}
where $\mathbf{B}_\text{tot} = \mathbf{B} + \mathbf{b}(t)$ is the total magnetic field, $\mathbf{B}$ is its constant component and $\mathbf{b}(t)$ is the time-dependent component induced by the rf coil, $\hat{\gamma}$ is the relaxation matrix reducing to $1/T_1$ for the $\mathbf{S}$ component along $\mathbf{B}_\text{tot}$ and to $1/T_2$ for the orthogonal component, and $\mathbf{S}_\text{st} = \chi \mathbf{B}_\text{tot}$ is the stationary spin polarization in the field $\mathbf{B}_\text{tot}$.

The solution of Eq.~\eqref{eq:Bloch} for constant $\mathbf{B}_\text{tot}$ is well known: the component of the initial spin $\mathbf{S}_0$ along $\mathbf{B}_\text{tot}$ tends to $\mathbf{S}_\text{st}$ as $\exp(-t/T_1)$, while the transverse component precesses with the Larmor frequency and decays as $\exp(-t/T_2)$.

(i) {\it In the first experimental configuration} [Fig.~\ref{fig:T1T2}(a)] both $\mathbf{B}$ and $\mathbf{b}$ are directed along the sample normal ($z$-axis). Using the rf coil we rapidly change the magnitude of the magnetic field. Initially (at $t<0$) $B_\text{tot} = B$, than $B_\text{tot}$ is rapidly increased, so that at $0 \leq t<T_\text{rf}$, $B_\text{tot} = B + b$, where $T_\text{rf}$ is the duration of the field pulse, and finally at $t \geq T_\text{rf}$ the magnetic field is reduced to $B$. In this case, the measured spin component is given by:
\begin{equation}
S_z(t) = \chi B + \chi b \times
\begin{cases}
   0,  t<0\\
   1 - \exp\left(-\frac{t}{T_1}\right), 0 \leq t<T_\text{rf} \\
   \left[\exp\left(\frac{T_\text{rf}}{T_1}\right) - 1\right]\exp\left(-\frac{t}{T_1}\right), t \geq T_\text{rf}.
\end{cases}
\end{equation}
Note, the the constant term $\chi B$ is not measured in the experiment due to the use of a synchronous detection scheme.

(ii) {\it In the second experimental configuration} [Fig.~\ref{fig:T1T2}(c)] $\mathbf{B}$ and $\mathbf{b}$ are directed perpendicular and along the sample normal, respectively. Using the rf coil we rapidly tilt the magnetic field from $\mathbf{B}$ to $\mathbf{B} + \mathbf{b}$ and after time $T_\text{rf}$ back to $\mathbf{B}$. In this case
\begin{equation}
S_z(t) \approx \chi b \times
\begin{cases}
   0,  t<0\\
   1 - \cos(\omega_\text{L} t)\exp\left(-\frac{t}{T_2^*}\right), 0 \leq t<T_\text{rf} \\
   \cos\left[\omega_\text{L} (t-T_\text{rf})\right]\exp\left(-\frac{t-T_\text{rf}}{T_2^*}\right), t \geq T_\text{rf},
\end{cases}
\end{equation}
where and in what follows we assume $b \ll B$, so that $\omega_\text{L} = |g|\mu_\text{B} B/\hslash$. Here, for simplicity, we also assume that the duration of the field pulse $T_\text{rf} \gg T_1$. Otherwise, the third expression should be multiplied by the constant $1 - \cos(\omega_\text{L} T_\text{rf})\exp\left(-T_\text{rf}/T_2^*\right)$.
Note that the magnetic field tilt equally affects all spins in the inhomogeneous ensemble. While the precession of each separate spin decays as $\exp(-t/T_2)$, averaging over all spins with the spread of Larmor frequencies $\Delta \omega_\text{L}$ results in a decay with the reduced time $T_2^*$, $1/T_2^*=1/T_2+\Delta \omega_\text{L}$.

(iii) {\it In the third experimental configuration} [Fig.~\ref{fig:T1T2}(e)] $\mathbf{B}$ and $\mathbf{b}$ are directed perpendicular and along the sample normal, respectively. We apply a sinusoidal voltage to the rf coil resulting in $\mathbf{b}(t) = \mathbf{b}^\text{a}\sin(\omega t)$. Here we consider only the resonant case $\omega = \omega_\text{L}$. Initially, at $t<0$, the spin polarization is in equilibrium and equals to $\mathbf{S}_\text{st} = \chi \mathbf{B}$. Nevertheless, we consider the more general case of an arbitrary initial spin polarization $S_0$ along $\mathbf{B}$, which can be also contributed by the optical pumping. The oscillating field $\mathbf{b}(t)$ drives a gradual declination of $\mathbf{S}$ from the direction of $\mathbf{B}$ and its precession about $\mathbf{B}$. The spin component along the sample normal can be expressed as
\begin{multline}
S_z(t) = \{\cos(\Omega_\text{R}t)\exp(-\frac{t}{\tau})+\\
+\left[\frac{1}{\tau\Omega_\text{R}}-\frac{|g|\mu_\text{B} b^a \eta}{2\hslash \Omega_\text{R}}\frac{S_0}{\chi B}\right]\sin(\Omega_\text{R}t)\exp(-\frac{t}{\tau})-1\} \times \\
\times \frac{\chi B}{\eta} \cos(\omega t),
\label{eq:SzRF}
\end{multline}
where
\begin{eqnarray}
\eta = \frac{T_1}{2} \frac{|g|\mu_\text{B} b^a}{\hslash}+\frac{2}{T_2} \frac{\hslash}{|g|\mu_\text{B} b^a}.
\end{eqnarray}
The first two terms describe Rabi oscillations with the frequency and decay time given by the following equations:
\begin{equation}
\Omega_\text{R} = \left[\left(\frac{g\mu_\text{B}b^a}{2\hslash}\right)^2+\left(\frac{1}{T_1}-\frac{1}{T_2}\right)^2\right]^{1/2},\\
\end{equation}
\begin{equation}
\tau = \frac{2T_1 T_2}{T_1 + T_2}.
\end{equation}
The Rabi oscillations were observed directly in the spin dynamics in the work \cite{Belykh2019}. The third term in Eq.~\eqref{eq:SzRF} gives the steady state spin precession. The amplitude of the spin precession at $t \gg \tau$ reaches its maximal value of $S_z^\text{max} = \sqrt{T_2/T_1}\chi B / 2$ at $b^a = 2\hslash/|g|\mu_\text{B} \sqrt{T_1 T_2}$, which is less than 1~mT for $T_1 \sim T_2 \sim 300$~ns. After the rf field is switched off at $t = T_\text{rf}\gg \tau$, the spin polarization decays as
\begin{equation}
S_z(t) = -\frac{\chi B}{\eta} \cos(\omega t) \exp\left(-\frac{t-T_\text{rf}}{T_2}\right).
\label{eq:rfDecay}
\end{equation}
For strongly inhomogeneous spin ensemble with $T_2^* \ll T_2$, the rf field affects only spins with $\omega_\text{L} - 1/T_2 \lesssim \omega \lesssim \omega_\text{L} + 1/T_2$ \cite{Belykh2019}, while the Larmor frequency spread of the entire ensemble is about $1/T_2^*$. Therefore, the decay time in Eq.~\eqref{eq:rfDecay} is not strongly affected by the inhomogeneous broadening and is reduced at most twice (being still much longer than $T_2^*$, e.g., for localized electrons). Nevertheless, the steady-state spin precession amplitude is reduced by a factor $T_2/T_2^*$ and its maximal value can be finally evaluated as
\begin{equation}
S_z^\text{max} = \frac{1}{2}\frac{T_2^*}{\sqrt{T_1 T_2}}\chi B.
\end{equation}

{\it The optical pumping} with circular polarization along $\mathbf{B}$, acting before the rf pulse arrival creates an initial spin polarization different from $\chi B$:
\begin{equation}
S_0^{\sigma \pm} = \chi B \mp \frac{\alpha T_1}{2 E_\text{las} l n_\text{e}} P_\text{OP},
\end{equation}
where $P_\text{OP}$ is the pumping power density, $E_\text{las}$ is the laser photon energy, $n_\text{e}$ is the density of resident electrons, $l$ is the penetration depth of the pump beam into the sample, and $\alpha$ is the spin orientation efficiency i.e. the probability to completely polarize the spin of a resident electron with one photon. $\alpha$ is related to the degree of circular polarization of the photoluminescence upon circularly polarized excitation, which was shown to be very low (a few percent) for $n$-GaAs in the metallic phase \cite{Dzhioev2002}. Note that the spin polarization is normalized to the total number of resident electrons. Taking into account the initial spin polarization $S_0 \neq \chi B$ in Eq.~\eqref{eq:SzRF}, we finally arrive at Eq.~\eqref{eq:SP} giving the spin precession amplitude at the end of the rf pulse, where $S^\text{a}$ and $\kappa$ can be easily found from Eq.~\eqref{eq:SzRF}.

\section*{Conclusion}
We have developed methods for spin manipulation and readout using combined rf and optical fields to enable measurement of the main spin relaxation times $T_1$, $T_2$ and $T_2^*$ using a radio-optical pump-probe technique. The three corresponding experimental configurations as well as their limitations are summarized in Table~\ref{Tab}. When measuring $T_1$ the signal is limited by the strength of the varying magnetic field, whereas there is no limitation for the constant magnetic field $B$. When measuring $T_2^*$, the signal is also limited by the varying magnetic field strength, while the rate of its variation which should be large compared to the Larmor frequency imposes a clear limitation on the constant magnetic filed. In the EPR-like experiments allowing to determine $T_2$ we lifted the limitations on the signal level using additional optical pumping along $\mathbf{B}$. Furthermore, we proposed a scheme allowing one to perform these experiments at high frequencies corresponding to high magnetic fields. This paves the way for an application of the radio-optical pump-probe technique to a wide range of material systems.  

\section*{Acknowledgments}
We are grateful to A.~R.~Korotneva for technical help. The work was supported by the Russian Science Foundation through grant No. 18-72-10073. High-frequency experiments were supported by the Deutsche Forschungsgemeinschaft in the frame of the ICRC TRR 160 (project A1).

\end{document}